\documentclass[iop]{emulateapj}
\usepackage{natbib}

\usepackage{wasysym}
\usepackage{txfonts}
\slugcomment{version \today}
\shorttitle{Flaring Patterns in Blazars}
\shortauthors{Paggi et al.}

\begin{document}

\title{{Flaring Patterns in Blazars}}

\author{A. Paggi}
\affil{Dipartimento di Fisica, Universit\`{a} di Roma ``Tor Vergata", Via della Ricerca Scientifica 1, I-00133 Roma, Italy}
\email{paggi@roma2.infn.it}

\author{A. Cavaliere}
\affil{Dipartimento di Fisica, Universit\`{a} di Roma ``Tor Vergata", Via della Ricerca Scientifica 1, I-00133 Roma, Italy\\
INFN - Sez. Roma 2, Via della Ricerca Scientifica 1, I-00133 Roma, Italy}

\author{V. Vittorini}
\affil{INAF/IASF-Roma, Via Fosso del Cavaliere 1, I-00100, Roma, Italy}

\author{F. D'Ammando}
\affil{INAF/IASF-Roma, Via Fosso del Cavaliere 1, I-00100, Roma, Italy\\
INAF/IASF Palermo, Via Ugo La Malfa 153, I-90146, Palermo, Italy}

\and

\author{M. Tavani}
\affil{Dipartimento di Fisica, Universit\`{a} di Roma ``Tor Vergata", Via della Ricerca Scientifica 1, I-00133 Roma, Italy\\
INAF/IASF-Roma, Via Fosso del Cavaliere 1, I-00100, Roma, Italy}

\begin{abstract}
Blazars radiate from relativistic jets launched by a supermassive black hole along our line of sight;
the subclass of FSRQs exhibits broad emission lines, {a telltale sign of a gas-rich environment and high accretion rate}, contrary to the other subclass of the BL Lacertae objects.
We show that this dichotomy of the sources in physical properties is enhanced in their flaring activity.
The BL Lac flares yielded spectral evidence of being driven by further acceleration of highly relativistic electrons in the jet.
Here we discuss spectral fits of multi-\(\lambda\) data concerning strong flares of the two flat spectrum radio quasars 3C 454.3 and 3C 279 recently detected in \(\gamma\) rays by the \textit{AGILE} and \textit{Fermi} satellites.
We find that optimal spectral fits are provided by external Compton radiation enhanced by increasing production of
thermal seed photons by growing accretion. We find {such} flares to trace patterns on the jet power - electron energy plane that diverge from those followed {by flaring} BL Lacs, and discuss why these occur.
\end{abstract}


\section{Introduction}

Blazars radiate from {narrow, relativistic} jets that blaze on us when closely aligned with
our line of sight. The jets are launched by a central supermassive black hole (SMBH, masses \(M_{\medbullet}\sim {10}^{8 - 9}M_{\astrosun}\)) with bulk Lorentz factors \(\Gamma\) up to several tens, and yield beamed non-thermal radiations with observed fluxes and frequencies enhanced by aberration and Doppler effects of Special Relativity \citep{bbr,konigl86,urry}.

The subclass of the BL Lac objects show {few} signs of radiation from an accretion disk, and weak or intermittent if any emission lines; {such conditions indicate} little surrounding gas, and low current accretion rates on the order of \(\dot{m}\lesssim {10}^{-2}\) in Eddington units \citep[e.g.,][]{ghisellini2009}. Their spectra are well represented as a smooth spectral energy distribution (SED) \(S_\nu=\nu F_\nu\), featuring two peaks: one in the range of {radio} - soft X-ray frequencies, due to synchrotron emission by highly relativistic electrons inside the jet; and another at hard X-ray or \(\gamma\)-ray energies, due to inverse Compton upscattering by the same electrons of {the} seed photons provided by the synchrotron emission itself (synchrotron self-Compton radiation, SSC; \citealt{jones,marscher,maraschi}).

The other blazar subclass is constituted by the Flat Spectrum Radio Quasars (FSRQs); besides the non-thermal peaks, they also feature the strong broad emission lines and high Big Blue Bump typical of quasars \citep[e.g.,][]{peterson}, which yield evidence of plenty surrounding gas associated to high current accretion rates {with} \(\dot{m}\sim 1\). Such {an} accretion provides seed photons from outside the jets, that drive an additional contribution to the non-thermal radiations by the external Compton scattering \citep[EC; see][]{dermer93,sikora}. {These} EC contributions often dominate the \(\gamma\)-ray outputs of FSRQs \citep{maraschi01}.

Both {source types} exhibit intermittent episodes of major brightening on timescales of days or even shorter, with substantial flux surges (``flares") particularly at \(\gamma\)-ray
energies, as recently observed for 3C 454.3 {by} \textit{AGILE}-GRID \citep{pacciani} and for 3C 279 by \textit{Fermi}-LAT \citep{abdo}.

Here we examine how the flares of the two subclasses differ, and relate their differences to specific physical processes in the sources. Flares in BL Lacs have already been studied in \citet{paggi1} and related to current acceleration of the emitting electrons; so in this paper we focus on flaring activity of FSRQs.

\section{Radiation: FSRQs versus BL Lacs}

The SEDs of {all} blazars include a basic SSC process.
The synchrotron component is produced {since} the jets contain a magnetic field \(B\), highly relativistic electrons with random Lorentz factors \(\gamma\) up to \({10}^6 - {10}^8\), in addition to non-relativistic protons with a common density \(n\) within a {comoving} size \(R\); all flow toward the observer with bulk Lorentz factors \(\Gamma\sim {10}-{20}\) (see \citealt{bott}; also \citealt{celotti}). {Interacting with} seed photons emitted inside or outside the jet, the same electrons also radiate by inverse Compton in the SSC or EC process, respectively.

 \begin{figure*}[t!]
\centering
\includegraphics[scale=0.6]{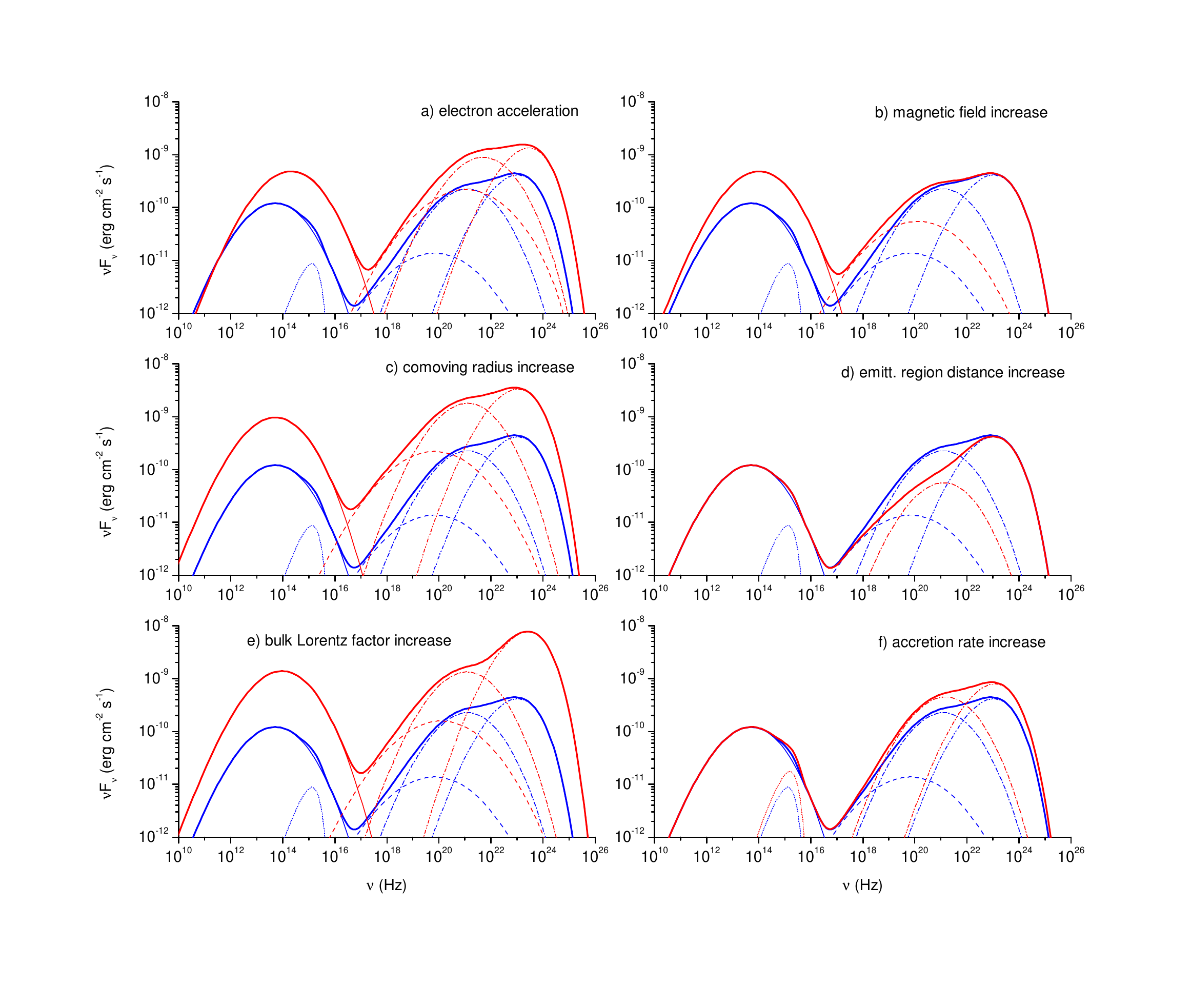}
\caption{Templates of the SEDs of a typical FSRQ. The pre-flare condition (blue lines) comprises the five spectral components with their parameters discussed in Sect. \ref{shapes}: synchrotron emission (full lines), SSC radiation (dashed lines), accretion disk emission (dotted lines), disk EC radiation (dot-dashed lines), and BLR EC radiation (double-dot-dashed lines). We also represent {flare conditions} (red lines), driven by different processes corresponding to an increase by a factor \(2\) in the key parameter: electron energy \(\gamma_p\) (panel a), magnetic field \(B\) (panel b), comoving radius \(R\) (panel c), height scale \(D\) (panel d), {bulk} Lorentz factor \(\Gamma\) (panel e), and accretion rate \(\dot{m}\) (panel f). In Table \ref{tabella} we give the detailed parameter values for the case (f), that is shown to be most relevant in Fig. \ref{3c279}.}
\label{examples}
 \end{figure*}

\subsection{{Spectra from SSC}}\label{seds}

We will mainly adopt for the electron energy distributions curved, log-parabolic shapes; adding to their value as fitting tools for {the associated spectral energy distributions, they are theoretically based as discussed in detail in Sect. \ref{discussion}.}

{In fact, our electron distributions will have the form}
\begin{equation}\label{logpar}
N(\gamma)=N_0\,
{\left({\frac{\gamma}{\gamma_0}}\right)}^{-s-r\log{\left({\frac{\gamma}{\gamma_0}}\right)}}\, .
\end{equation}
Here \(\gamma_0\) is the electron injection Lorentz factor and \(s\) is a constant contribution to the slope. {On a log-log plot, such distributions appear as parabolic humps} with curvature \(r\) \citep{massaro04}; {the parameters} may vary during flares.
The energetic content of such an electron population {is best} expressed in terms of the rms adimensional energy
\(\gamma_p={\left({{\int{\gamma^2 N{\left({\gamma}\right)}\, d\gamma}}/{\int{N{\left({\gamma}\right)}\, d\gamma}}}\right)}^{1/2}\); {this turns out to also provide the peak position for the function \(\gamma^2\,N(\gamma)\), that will enable us to compare BL Lacs and FSRQs energetics on a similar footing.}

{On the other hand, the SSC spectra radiated by such distributions are themselves curved \citep[and references therein]{massaro04,tramacere}, and their SEDs are effectively represented by the log-parabolic shapes}
\begin{equation}\label{logparsed}
S(\nu)=S_p{\left({\frac{\nu}{\nu_p}}\right)}^{-b\log{\left({\frac{\nu}{\nu_p}}\right)}}\, .
\end{equation}
{Here the} spectral curvature {is marked by \(b\), and the peak frequency \(\nu_p\) and flux \(S_p\)}; the two SED peaks closely yield the total {flux} in each component. For the synchrotron {component one has a peak frequency \(\nu_S \propto B\gamma_p^2\), a peak flux \(S_S\propto B^2 \gamma_p^2 n R^3\)}, and the spectral curvature \(b_s\approx r/5\). For the {IC component one has a peak frequency \(\nu_C\propto B\gamma_p^4\), a peak flux \(S_C\propto B^2 \gamma_p^4 n^2 R^4,\) and the spectral curvature \(b_C\approx r/10\)}, in the Thomson regime.

From an empirical standpoint, such curved spectra have been recognized to provide {very good spectral fits with uniform residuals}, from the X-ray \citep[e.g.,][]{landau,massaro04,tramacere,massaro08} to the sub-mm band \citep{gonzalez}.

\subsection{{Spectra from EC}}\label{shapes}

In the FSRQs in particular, the higher energy component usually features substantially larger fluxes with respect to the synchrotron component, {at variance with the yield of the SSC process in conditions of photon energy density not exceeding the magnetic one.}. This is often interpreted in terms of an additional EC component originated from seed photons emitted outside the jet, specifically by the broad line region (BLR) or by the inner accretion disk {\citep{dermer92,sikora,bottderm}}.

The disk emission is {widely} modeled as a sum of annular concentric surfaces radiating locally as a black body; their temperatures follow along the radius \(\mathcal{R}\) the profile \citep[see][]{frank},
\begin{equation}
T(r)=T_{*}\,{\left[{\frac{R_{i}^3}{\mathcal{R}^3}\left({1-\sqrt{\frac{R_{i}}{\mathcal{R}}}}\right)}\right]}^{1/4}\, ,
\end{equation}
where \(T_{*}\equiv{\left({3 G M_{\medbullet}\dot{M}}/{8 \pi {R_{i}}^3 \sigma}\right)}^{1/4}\), \(\dot{M}\) is the accretion rate and \(R_{i}\approx\, G M_{\medbullet}/c^2\) is the radius of the last stable orbit. The integrated emission has a peak flux \(S_d\propto \dot{M}\propto {T_{i}}^4\) at a peak frequency \(\nu_d\propto \dot{M}^{1/4}\propto T_{i}\). The BLR instead comprises a number of ``clouds" {intercepting some \(10\%\) of the radiation from the disk at distances \(\sim {10}^{17 - 18}\mbox{ cm}\), {with} ionization parameter values \(\sim {10}^{-1}\) \citep[see][]{peterson}}.

These {narrow spectra of} external photons {interact with} the electrons in the jet to yield EC radiation, with a SED again following shapes as represented by Eq. \ref{logparsed}. For the photons coming from the disk {to reach the emitting region at a height scale} \(D\), the EC radiation has a peak flux \(S_d\propto\dot{M}\,\gamma_p^2\,R^3\,n\,D^{-2}\,\Gamma^{-1}\), a peak frequency \(\nu_d\propto \dot{M}^{\frac{1}{4}}\,\gamma_p^2\,\Gamma^{-1}\), and the spectral curvature \(b_d\approx r\) close to the electron {distribution curvature}. For photons coming from the BLR we have a peak flux \(S_\mathcal{B}\propto \dot{M}\,\gamma_p^2\,R^3\,n\,\Gamma\),
a peak frequency \(\nu_\mathcal{B}\propto \dot{M}^{\frac{1}{4}}\,\gamma_p^2\,\Gamma\), {while} the spectral curvature is again \(b_\mathcal{B}\approx r\).

We recall that the observed frequencies are enhanced by the beaming factor \(\delta\), while the fluxes are boosted by \(\delta^4\), with \(\delta\approx 2\Gamma\) for small viewing angles \(\sim 1/\Gamma\) \citep{bbr}. {As mentioned above, the} EC flux is further {multiplied} by factors \(\Gamma^{-1}-\Gamma^2\) depending on the scattering geometry for the disk and the BLR photons, respectively \citep{dermer02,ghisellinitavecchio}.

Examples of these emissions and contributions to the SEDs are {numerically} computed and illustrated in Fig. \ref{examples}, for both the low and high states of a typical FSRQ with \(M_{\medbullet}\approx{10}^8 M_{\astrosun}\), \(\dot{m}\approx 1\), \(\Gamma\approx 10\), \(R\approx {10}^{17}\mbox{ cm}\), \(B\approx 1\mbox{ G}\) and \(n\approx {10}^2\mbox{ cm}^{-3}\), and \(\gamma_p\approx{10}^2\). These will be used in Sect. \ref{changes} as {a guide} to focus the main flare drivers, preliminary to our detailed fits in Fig. \ref{3c279}.

\section{Flares of 3C 454.3 and 3C 279}\label{sectflares}

The dominant process driving the blazar flares can be tested from the observed spectral changes compared with the templates provided by Fig. \ref{examples}.

\subsection{Spectral {variations}}\label{changes}

In several flaring BL Lacs the SEDs have been observed to increase both their peak heights and positions \citep[see][]{paggi1}; in particular, the synchrotron peak considerably shifting up in frequency constitutes clear evidence of electron \textit{acceleration}.

\begin{figure*}[t!]
\centering
\includegraphics[scale=0.5]{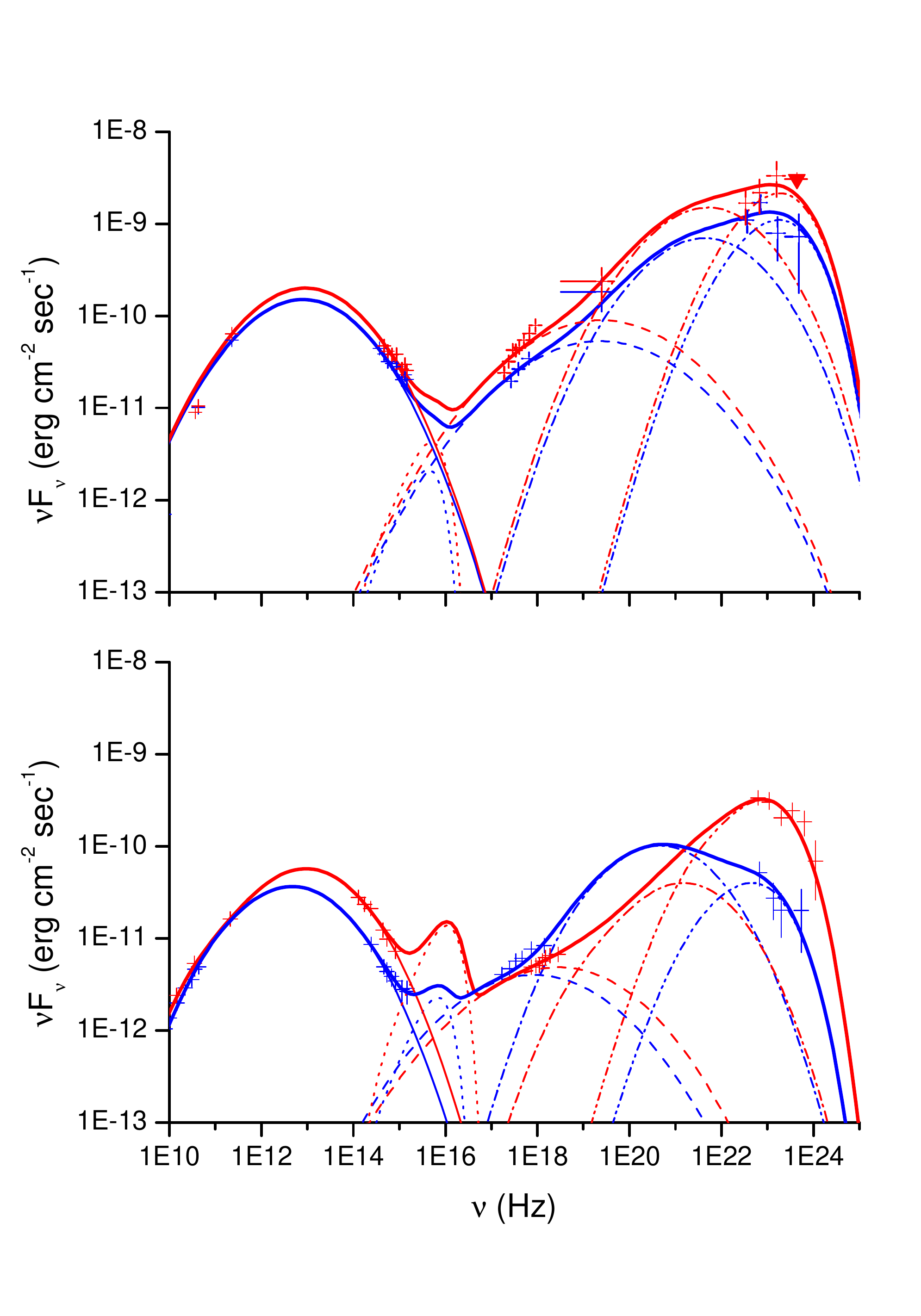}
\caption{Our fits to the SEDs of two specific FSRQs. ({Top} panel), low (blue lines) and high (red lines) state of the FSRQ 3C 454.3 as recently reported by \citet{pacciani}; ({bottom} panel), low and high states of the FSRQ 3C 279 \citep{abdo}. Model parameters are given in Table \ref{tabella}.}
\label{3c279}
 \end{figure*}

On the other hand, in several FSRQs the dominant and growing second spectral peak, along with the nearly stable synchrotron frequency, favor the EC process driven by {\textit{increasing} seed} photon flux. {This is} shown in Fig. \ref{examples}-f, {in comparison with} the alternatives presented in Fig. \ref{examples}-a, b, c,{ d, e}. Correspondingly, our fits to the specific SEDs of 3C 454.3 and 3C 279 are presented in Fig. \ref{3c279}, with model parameters given in Table \ref{tabella}. {Notice the larger particle densities found in FSRQs than in BL Lacs \citep{paggi2}, consistent with different matter contents in the surroundings of the two source kinds.}

We stress that for both sources the fits require considerably {\textit{larger}} values of \(\dot{m}\) in the flares, with limited variations of the other parameters; for 3C 279, however, the low X rays in the high state imply {also a larger distance} \(D\) of the emitting region. {Source transparency to pair production process via \(\gamma\)-\(\gamma\) interaction is assured by the optical depths \(\tau_{\gamma\gamma}= 0.03\) and \(\tau_{\gamma\gamma}= 0.24\) for \(h\nu\lesssim 1\) GeV, in the high state of 3C 454.3 and 3C 279, respectively.}

So the large flares of the FSRQs 3C 454.3 and 3C 279 appear to be dominated by thermal seed photons increasing their {fluxes much more than the frequencies}, as expected from the increasing \textit{accretion} rates discussed in Sect. \ref{seds}.
To wit, the EC flux grows strongly in \(\gamma\) rays while in the IR-optical bands little or no increase occurs for the synchrotron peak frequency (see Fig \ref{3c279}).

{On the other hand, the modest rise of the observed X-ray fluxes during flares constrains any increase of \(n\); so increasing seed photons apparently dominate the \(\gamma\)-ray surges.}

\subsection{Jet Power}\label{powers}

To further probe the physical processes driving a flare, an important quantity is provided by the overall energetics. The total jet power \(L_T\) is comprised of the radiative component \(L_{r}={\pi}\, D_L^2\, F/4{\Gamma^2}\), and of the kinetic powers in electrons and protons \(L_e+L_p={4}\pi\,R^2\,c^3\,n\,\Gamma^2\left({m_e\left\langle{\gamma}\right\rangle+m_p}\right)/3\), with a subdominant {magnetic component} \citep{celotti,paggi2}.

Here \(F\) is the flux observed at the luminosity distance\footnote{In the following, we adopt the standard flat cosmology with \(\Omega_\Lambda = 0.74\) and \(H_0 = 72 \mbox{ km}\mbox{ s}^{-1}\mbox{ Mpc}^{-1}\) \citep[{see}][]{dunkley}.} \(D_L(z)\), and \(m_e\) and \(m_p\) are the electron and proton masses. {One ``cold" proton per electron is assumed}, so for the FSRQs with average electron energies \(\left\langle{\gamma}\right\rangle < {10}^3\) the proton component to the energetics is considerable and hardly variable; this steadily feeds the energetics of the radio lobes \citep{celotti01}.

\subsection{{Flare evolutions}}\label{historical}

\begin{figure*}[t]
 \centering
 \includegraphics[scale=0.5]{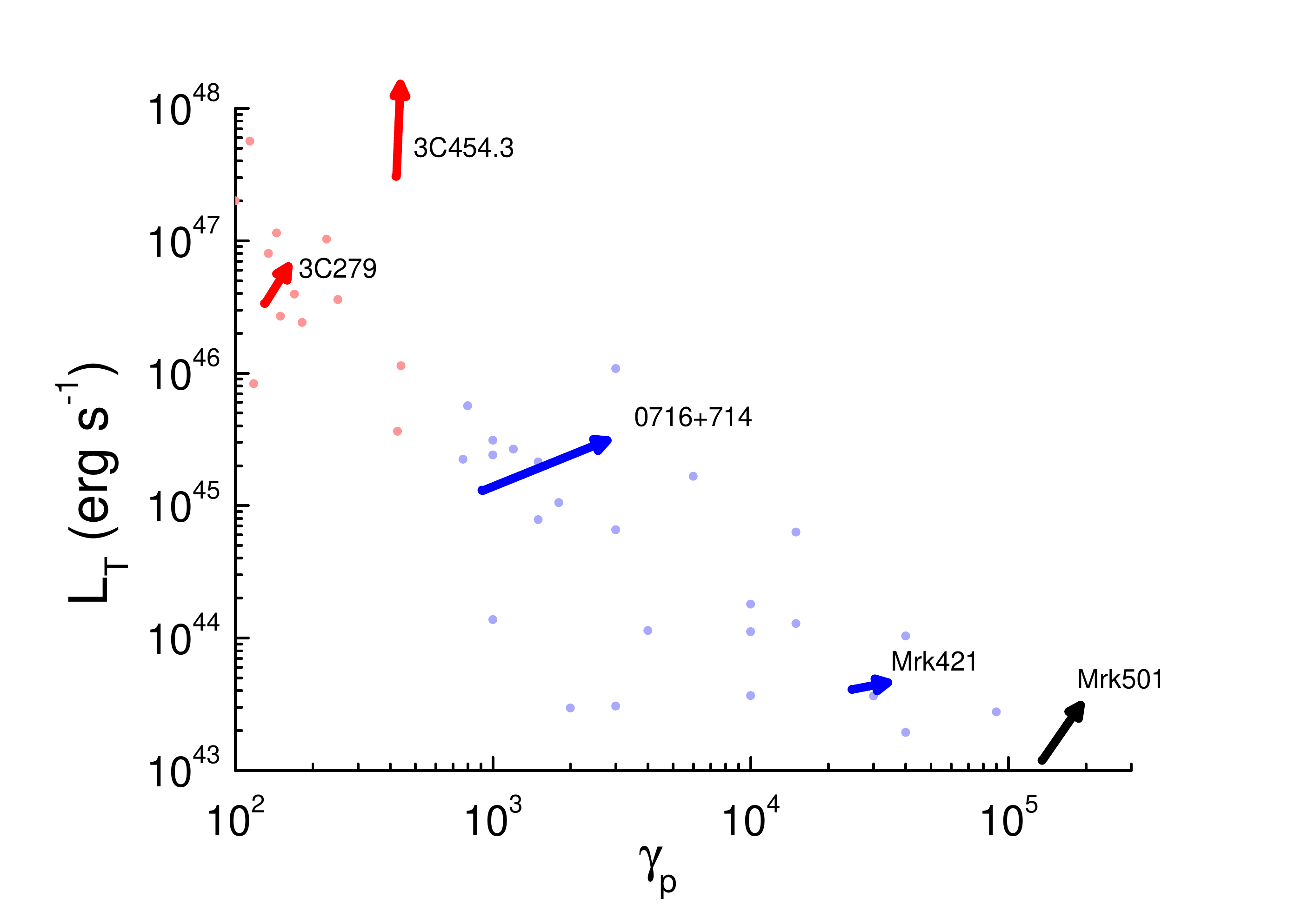}
\caption{Flaring patterns of blazars on the \(L_T\) - \(\gamma_{p}\) plane, superimposed to the bright blazar strip, covered by historical blazar states, {with data adapted from \citet{celotti} on taking for \(\gamma_p\) the peak of their \(\gamma^2 N(\gamma)\) distributions (see Sect. \ref{seds}).} Light blue dots represent BL Lacs and light red dots represent FSRQs. The arrows show the patterns of {evolution}: for gas-poor but luminous BL Lacs (blue arrows) the flare directions turn out to be substantially slanted to the right; for FSRQs (red arrows, extended by a factor 3 for better visibility, see Sect. \ref{historical}) we find closely vertical directions. {Our} actual high {and low} states are reported in detail by Paggi et al. (\citeyear{paggi2}) for BL Lacs, and in Table \ref{tabella} for the FSRQs. {The black arrow represents the flare pattern of the weaker and more massive Mrk501, not affected by constrains on luminosity or electron energy.}}
\label{flares}
 \end{figure*}

Our main {aim} here is constituted by {flares of} bright blazars; their historical conditions are effectively represented in the \(L_{T}\) - \(\gamma_{p}\) plane (see Fig.\ref{flares}), where they appear to be strung on average along a ``bright blazar strip" {\citep[see][]{cavalieredelia,bottderm,ghisellini02}}. The BL Lac sources reside in the {right-hand} region, {around the line where} the crossing time \(t_{cr}=R/c\) matches the {radiative} cooling time \(t_c\propto 1/\gamma w\) related to the energy density \(w\) in the magnetic or radiation field, to yield \(L_{T}\propto \gamma^{-1}\). On the other hand, the FSRQs lie in the {left-hand} region, {around the line} where the cooling time is matched by the acceleration time \(t_a\propto \gamma/E\) related to the effective electric field \(E\) (see \citealt{cavalieredelia}, \citealt{ghisellini2009}), to yield \(L_{T}\propto \gamma^{-2}\). The lower/left corner of the plane is populated by weaker sources out of our present scope \citep[see][]{padovani}.

We see from Fig. \ref{flares} that on this {plane} major flares {in our database move in directions different from the slope} of the bright blazar strip, into regions where the sources have higher luminosities and total powers and {faster cooling}. In particular, the FSRQs move \textit{vertically}, whereas luminous BL Lacs (e.g., S50716+714 and Mrk421) move in a substantially \emph{slanted} direction \citep{paggi2}. We interpret these patterns in terms of source structure changing as follows.

The flares of {gas-poor} BL Lacs are mainly driven by increasing \(\gamma\) causing {stronger} synchrotron radiation. {Cooling} is faster, but still can be balanced by shorter crossing times through a smaller source size (like is the case for jets with inner structure, see Tavecchio \& Ghisellini 2008, Giannios et al. 2009), to yield \(\gamma\sim 1/R\) and \(L_r \propto N/R^2\) in terms of the electron number \(N\propto n R^3\). Eventually, the source power may exceed the yield from {the feeble} current accretion; so it has to live on the alternative reservoir constituted by the the rotational energy of the SMBH, {that was} accumulated during previous massive accretion episodes, and {is now} tapped via the General Relativistic mechanism proposed by Blanford \& Znajek in \citeyear{bz}. This implies \(L\lesssim 8\times{10}^{45}\,M_{\medbullet}/{10}^9\,M_{\astrosun}\mbox{ erg/s}\); weaker and more massive sources like Mrk501 are not affected by such a constraint \citep[see also discussion by][]{paggi2}.

On the other hand, the flares of the luminous FSRQs are not so constrained in view of their high current accretion; the sources increase their luminosity \(L_r\propto \gamma^2 N S_d\), and in entering the fast cooling region are assisted by acceleration from \(E\) increasing with luminosity to yield \(\gamma\propto R\sqrt{E/L_r}\sim {const}\) \citep[see][]{cavaliere}. Note that \(L_T>L_r\) holds after the discussion in Sect. \ref{powers}, implying on the \(L_{T}\) - \(\gamma_{p}\) plane relatively short excursions {even during} substantial radiative flares; we have extended the relative arrows in Fig. \ref{flares} to highlight their directions, the relevant information in the present context.

{Thus in our synoptic view of Fig. \ref{flares} a relation is apparent between the source location and the flaring patterns of blazars on the \(L_T\) - \(\gamma\) plane.}

\section{{Discussion and conclusions}}\label{discussion}

Before concluding, we discuss why {such a} relation is not biased by our choice of log-parabolic shapes for the electron energy distributions and related SEDs given in Eqs. \ref{logpar} and \ref{logparsed}. We note that {cooling} (particularly in conditions of bright EC radiation) will erode them at high energies from their basic shapes. Actually, numerical simulations including synchrotron-type cooling have shown that these distributions retain an approximately log-parabolic shape around their peaks, with somewhat sharpened curvatures \citep[in particular their Fig. 4]{massaro06}.

\begin{table*}[!t]
\centering
\caption{Model parameters for 3C 454.3 and 3C 279 states described in the text; \(R\) is given in units of \({10}^{16}\mbox{ cm}\), \(n\) in units of \(\mbox{ cm}^{-3}\), \(B\) in \(\mbox{G}\), \(D\) in units of \({10}^{16}\mbox{ cm}\), \(L_r\) and \(L_T\) in \(\mbox{erg} \mbox{ s}^{-1}\).}\label{tabella}
\begin{tabular}{l|cccccccccc}
\hline
\hline
Source                & \(n\)   & \(\gamma_p\) & \(r\)   & \(R\)    & \(\Gamma\) & \(B\)   & \(D\)  & \(\dot{m}\) & \({L_r}\)                & \(L_T\)                \\
\hline
3C 454.3 (low state)  & \(5\)   & \(421.5\)    & \(0.9\) & \(71.9\) & \(16\)     & \(0.2\) & \(50.9\) & \(0.4\)     & \({1.4\times{10}^{47}}\) & \(3.1\times{10}^{47}\) \\
3C 454.3 (high state) & \(8\)   & \(425.3\)    & \(1.0\) & \(61.3\) & \(16\)     & \(0.2\) & \(51.2\) & \(1.0\)     & \({2.8\times{10}^{47}}\) & \(4.6\times{10}^{47}\) \\
\hline
3C 279 (low state)    & \(263\) & \(130.3\)    & \(1.1\) & \(4.7\)  & \(15\)     & \(1.3\) & \(4.4\)  & \(0.3\)     & \({3.5\times{10}^{45}}\) & \(3.4\times{10}^{46}\) \\
3C 279 (high state)   & \(140\) & \(137.3\)    & \(1.1\) & \(6.2\)  & \(15\)     & \(1.4\) & \(24.3\) & \(2.7\)     & \({6.4\times{10}^{45}}\) & \(3.9\times{10}^{46}\) \\
\hline
\hline
\end{tabular}
\end{table*}

The view is widened on using a Fokker-Planck equation that includes accelerations and slow cooling of an initial mono-energetic electron injection, as first proposed by \citet{kardashev} and recently computed in detail by Paggi et al. (\citeyear{paggi1}, and the refs. therein). Analytical solutions obtain in the form of a power series in the ratio \({t_a}/{t_c}\) of the acceleration \(t_a\) to the {cooling} time \(t_c\) defined in Sect. \ref{historical}.
The latter reads \(t_a/t_c =1.8\times{10}^{-15} \gamma^2 w/E\) in cgs units; on using the values of \(w\) obtained from our fits, this is found to be small out to \(\gamma\sim {10}^5\) for realistic values of the effective accelerating fields \(E\sim {10}^{-6}\, B\), in terms of the natural scale \(B\). Then the zeroth order solution is the log-parabola, while the first order correction steepens the energy distribution at the high end \(\gamma \gtrsim {10}^5\), mimicking a somewhat sharpened curvature that affects the SEDs at frequencies well beyond their peaks.

All that said, it is to be stressed that out to photon energies exceeding \(\approx 1\) GeV emitted by electrons with \(\gamma \gtrsim {10}^4\), broken power laws for the energy distribution (as adopted, e.g., by \citealt{donnarumma}, \citealt{vittorini}) yield closely similar shapes for the SEDs around their peaks; even less sensitive {is} the hump of \(\gamma^2 N(\gamma)\) also given by the {integral in} Sect. \ref{seds} that contributes the most to the energetics.
We conclude that the different {flare} spectra observed in the two blazar subclasses: BL Lacs and FSRQs, intrinsically relate to their \emph{different} physical properties during pre-flare conditions, i.e., to dearth or plenty of gas surrounding the central SMBH.

{In fact,} we have studied {in Sect. \ref{sectflares}} how this physical dichotomy is reflected in the flaring activity. We find that during the flares the BL Lac and the FSRQ follow different patterns, {indeed} \emph{diverging} ones on the \(L_T\) - \(\gamma_p\) plane.

We focused on the pattern derived from recent multi-\(\lambda\) observations of the two FSRQs, 3C 279 and 3C 454.3. We find {these} flares to {run} closely \textit{vertical} on the \(L_T\) - \(\gamma_p\) plane to be driven mainly by an increasing flux of the external photons, under further growth of the already substantial \textit{accretion} rates. Meanwhile the electron energies hardly increase, being constrained by the shortening cooling times in the denser radiative field.

These sources add to three BL Lacs that we previously studied (S50716+714, Mrk421 and Mrk501) and complete our picture. We interpret their flares in terms of \(\gamma_p\) increasing under further \emph{acceleration}, to yield patterns \textit{slanted} toward the \(\gamma\) axis; an effect reinforced by saturation of the power \(L_T\) due to the limited output extractable from a rotating SMBH via the Blanford-Znajek mechanism \citep{vittorini,paggi2}.

So we propose the following pattern for {the} blazar flares {analyzed here}: flaring spectral changes relate to the pre-flare \textit{positions} of BL Lacs and FSRQs on the \(L_T\) - \(\gamma_p\) plane, specifically, to their respective branch of the bright blazar strip; this is because the constraints that set such positions are retained or reinforced during the flares. To wit: bright sources lying in the upper branch of the strip flare up almost \textit{vertically} as expected from the cooling constraint to \(\gamma\), while sources in the lower right branch move almost \textit{horizontally} when luminous, as expected from the BZ constraint to output given their BH masses \(M_{\medbullet}\sim{10}^8\div {10}^9\,M_{\astrosun}\).

Such a picture may be tested on more sources with \textit{Fermi} and multi-wavelength data; in particular, it will be fruitful to study any interlopers between gas-rich powerful FSRQs and  gas-poor BL Lacs (that is, lower-luminosity FSRQs and the LBLs, low-peaked BL Lacs) during their flares on the \(L_T\) - \(\gamma_p\) plane in search of any divide or smooth rotation between these patterns. This may be the case for the BL Lac sources PKS 0537-441, AO 0235+164 and PKS 0426-380; with their weak broad lines and with \(\dot{m}\sim 0.1\) \citep[see][]{ghisellini2009} {they} may constitute transitional objects between FSRQs and BL Lacs. Further investigation of such objects will help understanding their nature and their stance in the above picture.

\acknowledgments
We acknowledge useful comments and suggestions by our anonymous referee. A. P. thanks the Harvard-Smithsonian Center for Astrophysics for hospitality, and F. Massaro in particular for useful discussions during completion of the present work.

\end{document}